\documentstyle[seceq,twocolumn,epsf]{jpsj}
\newcommand{\rmd}{{\rm d}}

\newcommand{\iu}{{\rm i}}

\def\runtitle{Kramer-Pesch effect in chiral p-wave superconductors}
\def\runauthor{Yusuke {\sc Kato} and Nobuhiko {\sc Hayashi}}

\title
{Kramer-Pesch effect in chiral p-wave superconductors}
\author
{Yusuke {\sc Kato}\footnote{E-mail: yusuke@phys.c.u-tokyo.ac.jp}
and
Nobuhiko {\sc Hayashi}$^{1,}$\footnote{E-mail: hayashi@mp.okayama-u.ac.jp}} 
\inst
{Department of Basic Science, University of Tokyo, Tokyo 153-8902\\
and\\
$^1$Computer Center, Okayama University, Tsushima-naka 3-1-1 Okayama 700-8530}

\recdate{\hspace{3cm}                                           }

\abst
{The pair-potential and current density around a single vortex of the two-dimensional chiral p-wave superconductor with ${\mib d}={\mib z}\left(p_x \pm \iu p_y\right)$ are determined self-consistently within the quasiclassical theory of superconductivity. Shrinking of the vortex core at low temperatures are considered numerically and analytically. Temperature-dependences of the spatial variation of pair-potential and circular current around the core and density of states at zero energy are the same as those in the isotropic s-wave case. When the senses of vorticity and chirality are opposite, however, we find two novel results; 1) the scattering rate due to non-magnetic impurities is considerably suppressed, compared to that in the s-wave vortex. From this observation, we expect that the chiral p-wave superconductors provide the best chance to observe the shrinking of the vortex (^^ ^^ Kramer-Pesch effect") experimentally. 2) The pair-potential of chiral p-wave superconductors inside vortex core recovers a {\it combined time-reversal-Gauge symmetry}, although this symmetry is broken in the region far from the vortex core. This local recovery of symmetry leads to the suppression of the impurity effect inside vortex core.}
\kword
{Superconductivity, vortex, quasiclassical theory, Kramer-Pesch effect, chiral superconductors}
\begin{document}
\sloppy
\maketitle
\section{Introduction}
The superconducting state in Sr$_2$RuO$_4$~\cite{Maeno} is expected to be unconventional; a Knight shift measurement in planar magnetic field suggests odd-parity pairing with ${\mib d}$ vector perpendicular to the RuO$_2$ plane~\cite{Ishida}. Muon spin relaxation experiment suggests spontaneous magnetization~\cite{Luke}; this means that time-reversal symmetry is broken spontaneously in the superconducting state.     
The simplest pairing symmetry which is consistent with the above two experimental results and the tetragonal crystal symmetry is given by the odd-parity pairing with ${\mib d}=\hat{\mib z}\left(p_x\pm\iu p_y\right)$~\cite{Sigrist}. This is commonly called the ^^ ^^ chiral p-wave superconducting state" because the Cooper pair has a definite angular momentum $L_z=\pm 1$ in the internal motion. A singly quantized vortex is, on the other hand, a configuration of the pair-potential where the center-of-mass motion of Cooper pair has an angular momentum. The interplay between vorticity and chirality would bring about intriguing phenomena in vortex physics of Sr$_2$RuO$_4$. Indeed, several authors~\cite{Volovik,Bocquet,Ivanov,Kato} have argued that the impurity effects in vortex core of chiral p-wave superconductor is considerably suppressed, compared to that in s-wave vortex core. The fact that the {\it superclean} regime is realized within the core of chiral p-wave vortex has several implications. First, the flux flow conductivity is expected to be enhanced at low temperatures in the chiral p-wave vortex states~\cite{Kato}. Second, a large Hall effect is also expected. Third, the shrinkage of vortex cores at low temperatures, which is referred to as ^^ ^^ the Kramer-Pesch (KP) effect"~\cite{KramerPesch}, is free from the smearing due to impurities. The vortex shrinking effect in chiral p-wave superconductors is the main issue in this paper. 

The KP effect, originally predicted for s-wave vortices, was studied numerically by Gygi and Schl\" uter~\cite{Gygi}, who gave a self-consistent solution to 
the Bogoliubov-deGennes (BdG) equation. More general cases were discussed by Volovik\cite{Volovik2}. Vortex shrinking in d-wave case was confirmed numerically by Ichioka {\it et al}. \cite{Ichioka}. Experimental studies on the vortex shrinking in NbSe$_2$ and cuprates have been reviewed recently by Sonier {\it et al}~\cite{Sonier}. 
For the chiral p-wave case, the electronic structure and charging effect were studied \cite{HeebThesis, M&H, Takigawa} within the framework of the Bogoliubov-deGennes (BdG) equation. While the analysis in the quantum limit $k_{\rm F}\xi\approx 1$ brings us the understanding of physics at extremely low temperatures, where the quantum effects becomes important\cite{MKM, HIIM, KM, F&T}, the quasiclassical condition $k_{\rm F}\xi \gg 1$ is satisfied in Sr$_2$RuO$_4$ with a $\xi_0$ (the coherence length at low temperature) $\approx 660$\AA~\cite{Akima}. 
In this paper, we will study the structure of chiral p-wave vortices by adopting a more appropriate framework; the quasiclassical theory of superconductivity~\cite{Eilenberger,LO68,Eliashberg,SereneRainer}. The quasiclassical scheme has previously been applied to the study of the pair-potential and the local density of states near domain wall\cite{Matsumotodomain}.  

 In the next section, we summarize the formulation of the quasiclassical theory of superconductivity. In \S~\ref{sec:Analytical Results}, we calculate the pair-potential analytically, only considering contributions to the gap equation from the chiral branch. In \S~\ref{sec:Numerical Results}, we present our numerical results for the pair-potential and establish the validity of our analytical findings of \S~\ref{sec:Analytical Results}. In \S~\ref{sec:Discussion}, we discuss aspects different from s-wave vortex cores. In \S~\ref{sec:Conclusion}, we summarize our results.        
\section{Formulation}\label{sec:Formulation}
\subsection{Quasiclassical theory of superconductivity}
We consider a two-dimensional superconductor with the circular Fermi surface. For simplicity, we consider the type II limit $(\kappa\equiv \lambda/\xi\rightarrow \infty)$ in this paper except in subsection~\ref{subsec:finitekappa}. In Sr$_2$RuO$_4$, the coherence length $\xi_0$ at zero temperature is expected to be $\approx 660$\AA~\cite{Akima} and much larger than the Fermi wave length, which is of the order of several angstroms. Since the atomic-scale structures are irrelevant in determining the pair-potential, we can factor out the rapidly oscillating terms which enter the wavefunctions of the BdG equation. The resulting equation (Andreev equation)~\cite{Andreev} is equivalent to the quasiclassical equation of superconductivity~\cite{Eilenberger,LO68,Eliashberg,SereneRainer}, which we now explain briefly. The quasiclassical theory of superconductivity is described in the equilibrium case in terms of the quasiclassical green function
\begin{equation}
\hat g(z,{\mib r},\hat {\mib p})=\hat g(z,{\mib r},\alpha)
=\left(
\begin{array}{rc}
g&f\\
-\tilde f&-g
\end{array}\right),
\label{greenfunction}
\end{equation}
which is a $2\times 2$ matrix in particle-hole space and is a function of (complex) frequency $z$, the direction $\hat{\mib p}=(\cos\alpha,\sin\alpha)$ of momentum ${\mib p}=p_{\rm F}\hat{\mib p}$, a point ${\mib r}=r(\cos \phi,\sin \phi)$ in real space. The equation of motion for $\hat g$ is given by
\begin{equation}
-\iu  v(\hat{\mib p})\cdot{\mib\nabla} \hat g=\left[\epsilon \hat\tau_3-\hat \Delta({\mib r},\hat {\mib p}),\hat g\right],
\label{qclequation}
\end{equation}
supplemented by the normalization condition 
\begin{equation}
\hat g^2=-\pi^2 \hat 1, 
\label{normalization}
\end{equation}
where $\hat 1$ is the $2\times 2$ unit matrix. 

Here $\hat \tau_3$ is a Pauli matrix,
\begin{equation}
\left(
\begin{array}{rc}
1&0\\
0&-1
\end{array}\right)
\label{pauli3}
\end{equation}
and $\hat \Delta$ is given by
\begin{equation}
\left(
\begin{array}{rc}
0&\Delta({\mib r},\hat{\mib p})\\
-\Delta^*({\mib r},\hat{\mib p})&0
\end{array}\right), 
\label{deltahat}
\end{equation}
where $\Delta({\mib r},\hat{\mib p})$ is the pair-potential. 

To determine the pair-potential self-consistently, we have to consider the case where $z$ is a Matsubara frequencies $\iu \omega_{m}=\iu \pi T\left(2m+1\right)$. Setting $z=\epsilon+\iu\delta$ with real $\epsilon$, on the other hand, we can obtain the density of states from $-{\rm Im}g(\epsilon+\iu \delta)$.  

The self-consistent equation for the pair-potential 
\begin{equation}
\Delta({\mib r},\alpha)\equiv\Delta^+\left({\mib r}\right)e^{\iu \alpha}+
\Delta^-\left({\mib r}\right)e^{-\iu \alpha}
\label{Deltaralpha}
\end{equation}
is given by 
\begin{equation}
\Delta^\pm\left({\mib r}\right)=TV_p\sum_{|\omega_m|<\omega_{\rm c}}\int_{0}^{2\pi}\frac{\rmd \alpha}{2\pi} e^{\mp\iu \alpha}f(\iu\omega_m,{\mib r},\alpha).
\label{gapequation}
\end{equation}
Here the strength $V_p$ of the attractive interaction is given by
$$
V_p^{-1}\equiv\ln\left(T/T_{\rm c}\right)+\sum_{m=0}^{\omega_{\rm c}/(2\pi T)}\frac{1}{m+1/2}. 
$$

The current density ${\mib j}({\mib r})$ is given by
\begin{equation}
2N_0 e vT\sum_{|\omega_m|<\omega_{\rm c}} \int_0^{2\pi}\frac{\rmd \alpha}{2\pi}\hat{\mib p}g\left(\iu \omega_m,{\mib r},\alpha\right), 
\label{currentdensity}
\end{equation}
where $N_0$ denotes the normal state density of states and $e(<0)$ charge unit. 
The set of equations (\ref{greenfunction})-(\ref{gapequation}) possess two continuous symmetries, the rotational symmetry around $z$-axis with the generator $\hat L_z$ and U(1) Gauge symmetry with $\hat I$. The actions of these generators are written, respectively, by 
\begin{equation}
\hat L_z =-\iu\left(\partial/\partial \phi+\partial/\partial \alpha\right)
\label{generatorL} 
\end{equation}
and 
\begin{equation}
\begin{array}{ccc}
\hat I \Delta=\Delta,& \hat I \Delta^*=-\Delta^*,& \hat I g=0\nonumber\\ 
\hat I f=f,& \hat I \tilde f=-\tilde f. &  
\end{array}
\label{generatorU}
\end{equation}
We note that the expression (\ref{generatorL}) consists of the center-of-mass rotation ($-\iu \partial/\partial \phi$) and the internal rotations of the Cooper pairs ($-\iu \partial /\partial \alpha$). 

We consider two configurations of a single vortex with the following asymptotic forms for the pair-potential:  
\begin{equation}
\Delta({\mib r}\rightarrow \infty,\alpha)\rightarrow \left\{
\begin{array}{cc}
\Delta_\infty e^{\iu \left(\phi-\alpha\right)},&\quad\mbox{Case I}\\
\Delta_\infty e^{\iu \left(\phi+\alpha\right)},&\quad\mbox{Case II}
\end{array}\right.
\label{asymptoticsDelta}
\end{equation}
and for the green function:
\begin{equation}
\hat g(z,{\mib r},\alpha)\rightarrow \frac{\pi}{\sqrt{\Delta({\mib r},\alpha)^2-z^2}}
\left(
\begin{array}{cc}
-z,&\Delta({\mib r},\alpha)\\
-\Delta^*({\mib r},\alpha),&z
\end{array}
\right).
\label{asymptoticsgreen}
\end{equation}
In (\ref{asymptoticsDelta}), $\Delta_\infty$ denotes the amplitude of the pair-potential in the bulk and is given by
$$
1=\pi T V_p\sum_{|\omega_m|<\omega_{\rm c}}\left(\Delta_\infty^2+\omega_m^2\right)^{-1/2}.
$$. 

The asymptotic forms (\ref{asymptoticsDelta}) and (\ref{asymptoticsgreen}) determine the symmetries of the two vortices; each symmetry is described by the modified generator $\hat Q=\hat L_z -m \hat I$ with $m=0$ for case I and $m=2$ for case II. 

By solving $\hat Q\Delta({\mib r},\alpha)=0$, the general expression for pair-potential in case I is obtained as
\begin{equation}
\Delta({\mib r},\alpha)=\Delta_0^+(r)e^{\iu\left(-\phi+\alpha\right)}+
\Delta_0^-(r)e^{\iu\left(\phi-\alpha\right)}
\label{caseIDelta}
\end{equation}
and that in case II is given by
\begin{equation}
\Delta({\mib r},\alpha)=\Delta_0^+(r)e^{\iu\left(\phi+\alpha\right)}+
\Delta_0^-(r)e^{\iu\left(3\phi-\alpha\right)}. 
\label{caseIIDelta}
\end{equation}
Here $\Delta^\pm_0(r)$ denote functions of $r=|{\mib r}|$. In the next section, we will numerically determine the self-consistent solution for $\Delta^\pm_0(r)$.   
\subsection{Riccati Formalism}

In order to determine the pair-potential, we could solve the quasiclassical equation (\ref{qclequation}) and the gap equation (\ref{gapequation}). Instead of directly attacking eq.~(\ref{qclequation}), we reformulate the original quasiclassical equation (\ref{qclequation}) into Riccati-type equations for auxiliary functions\cite{Nagato1,Higashitani,Nagato2,SchopohlMaki,Schopohl,Eschrig}.
Let $\gamma(z,{\mib r},\alpha)$ and $\gamma^\dagger(z,{\mib r},\alpha)$ be the solutions of the following Riccati differential equations: 
\begin{subeqnarray}
\iu {\mib v}\cdot{\mib \nabla}\gamma=-\Delta^*\gamma^2-2\epsilon\gamma-\Delta\\
\iu {\mib v}\cdot{\mib \nabla}\gamma^\dagger=-\Delta\gamma^{\dagger 2}+2\epsilon\gamma^\dagger-\Delta^*. 
\label{Riccati1}
\end{subeqnarray}
The solution $\hat g$ of (\ref{qclequation}) can then be given by
\begin{equation}
\hat g=\frac{-\iu\pi}{1+\gamma\gamma^\dagger}
\left(
\begin{array}{cc}
1-\gamma\gamma^\dagger&2\gamma\\
2\gamma^\dagger&-1+\gamma\gamma^\dagger
\end{array}\right).
\label{twobytwo}
\end{equation}

In this parameterization, the normalization condition (\ref{normalization}) is automatically satisfied. 

For a given $\alpha$, the framing of space 
$$
{\mib r}=s \hat p + b \hat z\times \hat p
$$ 
is naturally introduced. Since the Fermi surface is circular symmetric, ${\mib  v(\hat {\mib p})}\cdot {\mib \nabla}$ can be rewritten as
$$
v\hat{\mib p}\cdot{\mib \nabla}=v\partial/\partial s.
$$
Thus, eq. (\ref{Riccati1}) reduces to a one-dimensional problem on a straight line (quasiclassical trajectory) with a constant impact parameter $b=r\sin(\phi-\alpha)$. Equations (\ref{Riccati1}) are to be solved, under the initial conditions:
\begin{subeqnarray}
\gamma=\frac{\Delta}{-\epsilon-\iu \delta-\iu\sqrt{\left|\Delta\right|^2-\epsilon^2}}, \quad\mbox{ for }s=-s_{\rm c}\\
\gamma^\dagger=\frac{\Delta^*}{\epsilon+\iu \delta+\iu\sqrt{\left|\Delta\right|^2-\epsilon^2}}, \quad\mbox{ for }s=s_{\rm c},
\label{ic1}
\end{subeqnarray}
where $s_{\rm c}$ is a positive real cut-off. 
Numerical calculation of $\gamma$ for ${\rm Im}z>0$ is stable if (\ref{Riccati1}~a) is integrated starting with (\ref{ic1}~a) towards increasing $s$. The calculation of $\gamma\dagger$ for ${\rm Im}z>0$, on the other hand, is numerically stable if eq. (\ref{Riccati1}~b) with the initial condition (\ref{ic1}~b) is solved with decreasing $s$\cite{Schopohl}. Obviously eq. (\ref{gapequation}) contains $f(\iu \omega_m,{\mib r},\alpha)$ with negative $\omega_m$. Such contributions can, however, be eliminated easily with the use of the symmetry relation $f(\iu \omega_m,{\mib r},\alpha)=-f(-\iu \omega_m,{\mib r},\alpha+\pi)$\cite{Matsumotodomain}. 

Numerical integrations of (\ref{Riccati1}) are performed by the adaptive stepsize control Runge-Kutta method\cite{Recipe}. The cut-off frequency $\omega_{\rm c}$ is taken as $10\Delta_\infty$, where $\Delta_\infty$ is the modulus of the pair-potential away from vortex center at zero temperature. The cut-off length $s_{\rm c}$ is taken as $3\xi_0,6\xi_0,12\xi_0$ ($\xi_0\equiv v/\Delta_\infty$). The choice of cut-off depends on temperatures. 

\section{Analytical Results}\label{sec:Analytical Results} 
\subsection{Chiral branch and Kramer-Pesch effect}
In both cases (I and II), each vortex has a zero-energy bound state for trajectory with $b=0$, because the order parameter is real up to an overall phase and changes its sign when going from $s=-\infty$ to $s=\infty$\cite{Atiyah}. On a trajectory with nonzero impact parameter, there is at least one Andreev bound state with nonzero energy. The lowest bound states for each impact parameter form a single branch, which corresponds to the Caroli-deGennes-Matricon mode\cite{Caroli}. Roughly speaking, this branch has energy linear in the impact parameter $b$ (^^ ^^ chiral branch "in this sense) and merges into the bulk continuum spectrum with $b\sim \xi_0$. Therefore, the dispersion can be estimated as $E\sim \Delta_\infty b/\xi_0$. This chiral branch plays an important role in the self-consistent calculation of the pair-potential, as explained below.  

Overall features of the pair-potential around a single vortex at a wide range of temperatures lower than $T_{\rm c}$ can be described by the GL theory; Although the GL theory is valid only near $T_{\rm c}$ in principle, the length scale which the bulk quasiparticle introduces into the gap equation (\ref{gapequation}) is of the order of $\xi_0$ even away from $T_{\rm c}$. However, at temperature $T (\ll T_{\rm c})$, the chiral branch with energy comparable to $T$ gives a large contribution to the pair-potential in the gap equation (\ref{gapequation}) at the radius $r\sim \xi_1=T\xi_0/\Delta_\infty$ from the vortex center. Thus, the chiral branch introduces {\it another length} which scales the spatial variation of pair-potential in the vicinity of vortex center. The emergence of $\xi_1$ at low temperatures can be regarded as a result of the shrinking of vortex core (KP effect\cite{KramerPesch}). 
  
For the chiral branch with $|z|\ll \Delta_\infty$, the green function $\hat g$ can be obtained analytically\cite{KramerPesch,Eschrig,Kato}. The contribution in the gap equation (\ref{gapequation}) from the chiral branch can, therefore, be calculated analytically. Further we assume that near the vortex center, contributions in (\ref{gapequation}) to the pair-potential from other excitations are negligible. Within this assumption, the self-consistent determination of the pair-potential becomes feasible. This assumption will be justified by the numerical calculation in \S~\ref{sec:Numerical Results}.

\subsection{Case I: $\Delta({\mib r},\alpha)\rightarrow \Delta_\infty e^{\iu \left(\phi-\alpha\right)}\label{subsec:CaseI}
$}
The analytic expression for the green function of the chiral branch in the s-wave case was first obtained in ref.~(\citen{KramerPesch}). We, however, follow the method presented in ref.~(\citen{Eschrig}), which is based on the Riccati formalism. The outline of the calculation of $\hat g$ is as follows: First, calculate $\gamma$ and $\gamma^\dagger$ for $z=0$ and $b=0$. This nonlinear differential equation happens to be elementarily integrable. Second, calculate $\gamma$ and $\gamma^\dagger$ up to linear order in $z$ and $b$. The problem reduces to a first-order linear differential equation, which can be easily solved. Third, substitute the resulting $\gamma$ and $\gamma^\dagger$ into (\ref{twobytwo}). The zero-th order with respect to $z$ and $b$ is cancelled in the denominator in (\ref{twobytwo}). Therefore, the first order term should be kept there. In the matrix part in (\ref{twobytwo}), on the other hand, zero-th order term with $z$ and $b$ survives and hence the first order contribution can be neglected. One can refer to refs.~(\citen{Kato}) and (\citen{Eschrig}), for the details of the method. 

\begin{figure}
\begin{center}
\epsfysize=150pt
\epsfbox{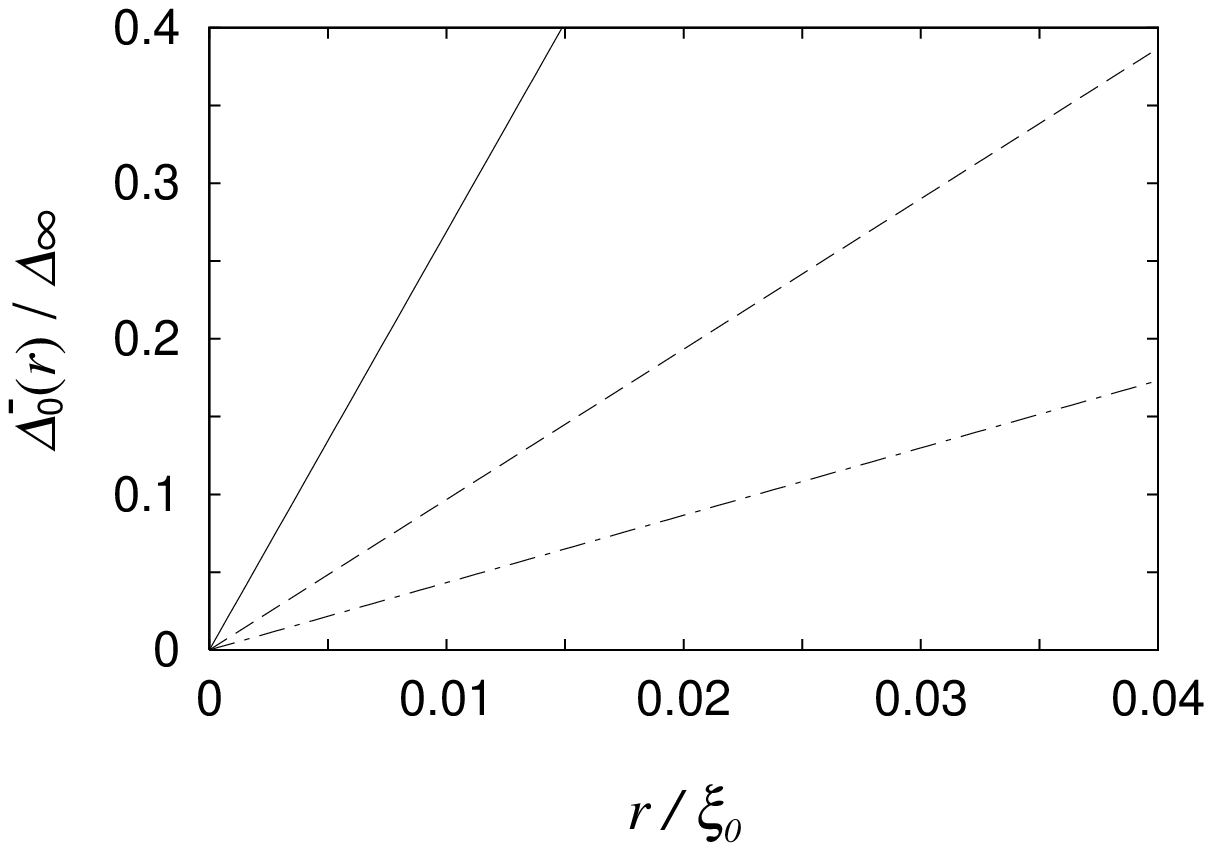}
\end{center}
\begin{center}
\epsfysize=150pt
\epsfbox{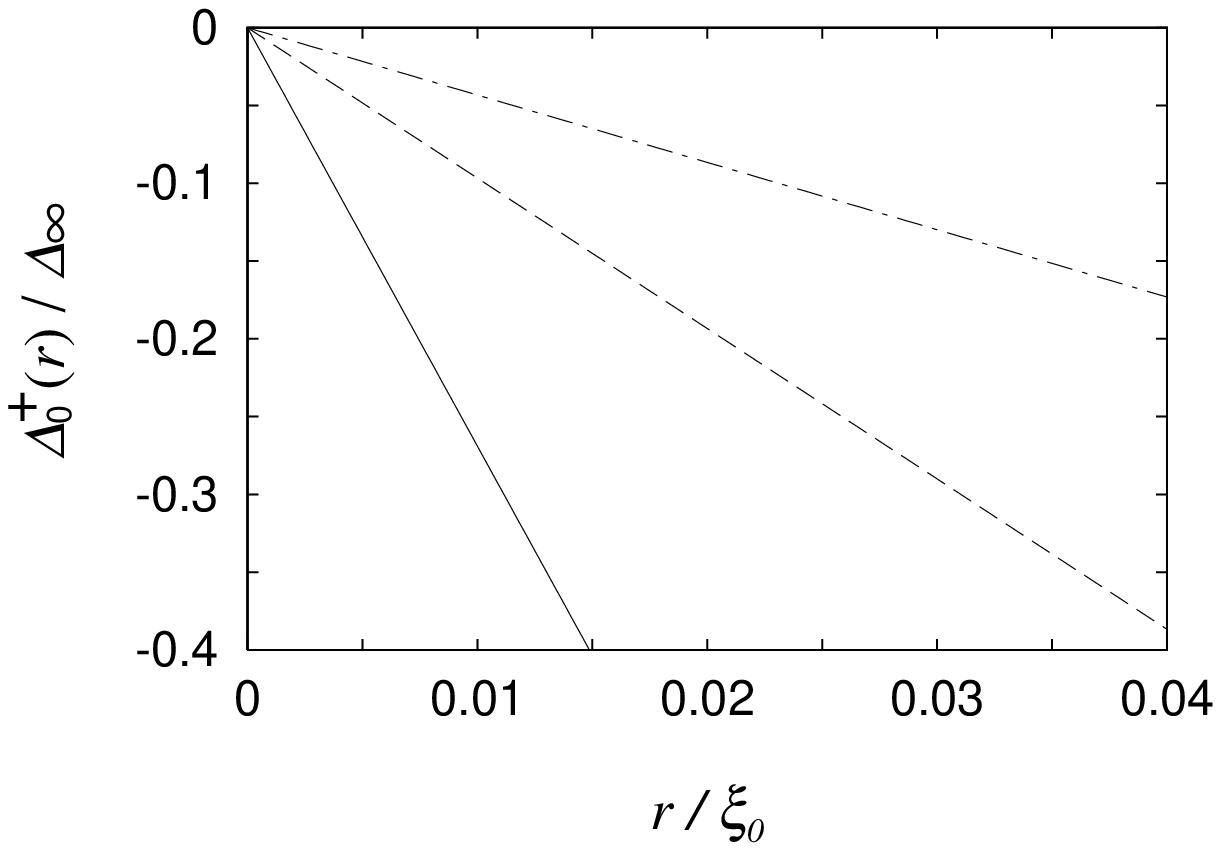}
\end{center}
\caption{Analytical results (eq.~(\ref{DeltalimitKP1}) in the text) for case I where total angular momentum is zero. 
The initial slopes of the pair-potentials (a) $\Delta_0^-(r)$ for the dominant component with negative chirality and (b) the induced component $\Delta_0^+(r)$ with positive chirality are shown as functions of the distance $r$ from the vortex center, for temperatures $T/T_c=0.02$ (solid line), 0.05 (dashed line) and 0.1 (dotted-dashed line). The vertical and horizontal axes are scaled, respectively, by the modulus of the bulk pair-potential $\Delta_\infty$ and the coherence length $\xi_0$ at zero temperature. }
\label{fig:1}
%
%
%
%
%
\end{figure}

The resulting expression for $\hat g(z,{\mib r},\alpha)$ is given by
\begin{equation}
\hat g\sim  \frac{\pi v \exp\left[-u_-(s)\right]\hat M_-}{2C_-\left(z-E_-(b)\right)},
\label{analytic1}
\end{equation}
with 
$$
\hat M_-=\left(
\begin{array}{cc}
1,&-\iu\\
-\iu,&-1
\end{array}
\right).
$$
Here $\exp\left[-u_-(s)\right]$ with
$$
u_-(s)=\frac{2}{v}\int_0^{|s|}\rmd s'\left\{\Delta_0^-(s')+\Delta_0^+(s')\right\}
$$
is essentially the spectral weight (up to a constant factor ) of the Andreev bound state for the quasiclassical trajectory specified by $b$ and $\alpha$. The function $\exp\left[-u_-(s)\right]$ takes the value of the order of unity for $s\ll \xi_0$ and much smaller than unity for $s\gg \xi_0$.  The constant $C_-$ in the denominator of (\ref{analytic1}) is defined by
\begin{equation}
C_-=\int_0^\infty\rmd s \exp\left[-u_-(s)\right].
\label{Cminus}
\end{equation}
The  symbol $E_-(b)$ denotes the energy of the Andreev bound state, which is given by
\begin{equation}
E_-(b)=\frac{b}{C_-}\int_0^\infty\frac{\rmd s}{s}\exp\left[-u_-(s)\right]\left\{\Delta_0^-(s')-\Delta_0^+(s')\right\}.
\label{Eminus}
\end{equation}
Substituting the expression for $f(z,{\mib r},\alpha)$ in (\ref{analytic1}) into (\ref{gapequation}), we obtain 
\begin{equation}
\Delta^\pm_0(r)=\frac{\mp \pi v V_p I_1(E_-'r/(2T))}{4C_-},
\label{DeltaKP1}
\end{equation}
with 
\begin{equation}
I_n(R)\equiv\int_0^\pi\frac{\rmd \alpha'}{\pi}\sin n\alpha' \tanh\left(R\sin\alpha'\right)\quad n=1,3,\cdots. 
\end{equation}
Here $E_-'$ is defined by $\rmd E_-(b)/\rmd b$. For $r\ll \xi_{1-}\equiv 2T/E'_-$, the expression (\ref{DeltaKP1}) reduces to 
\begin{equation}
\Delta^\pm_0(r)\rightarrow \frac{\mp \pi v V_p}{8 C_-}\left(\frac{r}{\xi_{1-}}\right). 
\label{DeltalimitKP1}
\end{equation}
From eq.~(\ref{DeltalimitKP1}), we can see that both components $\Delta^+_0$ and $\Delta^-_0$ vanish linearly with respect to the distance $r$ from the vortex center. Further, both components have the same amplitude. The relative phase between them is $\pi$. The initial slope of $\Delta^\pm_0(r)$ is shown in Fig.~\ref{fig:1}. 

The  length $\xi_{1-}$ can be estimated in the following way. 
In (\ref{Cminus}) and (\ref{Eminus}), $\exp\left[-u_-(s)\right]$ becomes unity for $s\ll \xi_0$ and much smaller than unity for $s\gg \xi_0$. Thus $C_-$ in (\ref{Cminus}) is of the order of $\xi_0$ and then $E'_-$ is reduced to 
$$
E'_-\sim \frac{1}{C_-}\int_0^{\xi_0}\frac{\rmd s}{s}\left\{\Delta_0^-(s')-\Delta_0^+(s')\right\}\sim \frac{\Delta_\infty}{\xi_0}\ln\left(\frac{\xi_0}{\xi_{1-}}\right).
$$
From the above expression, we obtain 
$$
\xi_{1-}/\xi_0=2T/(\xi_0 E'_-) \sim \left(T/\Delta_\infty\right)/\ln\left(\xi_0/\xi_{1-}\right),   
$$
which gives $\xi_{1-}/\xi_0\sim \left(T/\Delta_\infty\right)/\ln\left(\Delta_\infty/T\right)$. According to this argument, the slope of the pair-potential around the vortex center is found to diverge in the limit of zero temperature. 

The current density coming from the chiral branch (\ref{analytic1}) is given by ${\mib j}_-({\mib r})=j_-(r)\hat{\mib \phi}$ with 
\begin{eqnarray}
j_-(r)=\frac{\pi e v^2 N_0I_1\left(r/\xi_{1-}\right)}{2C_-}\sim \frac{\pi e v^2 N_0}{4C_-}\left(\frac{r}{\xi_{1-}}\right)\nonumber\\
\sim -j_{\rm c}(0)\left(\frac{r}{\xi_{1-}}\right). 
\label{jminus}
\end{eqnarray} 
Here $j_{\rm c}\equiv 2|e|N_0 v\Delta_\infty$ denotes the critical current in a homogeneous state at zero temperature. 
The contribution to the current density from the extended quasiparticles are expected to be of the order of $j_{\rm c}(r/\xi_0)$. At temperatures much lower than $T_{\rm c}$, where $\xi_{1-}\ll \xi_0$, the distribution of the current density becomes more concentrated around the vortex center. 

The shrinkage of the vortex core also affects the density of states (integrated over a vortex core):
\begin{equation}
N(\epsilon=0)\equiv-\frac{N_0}{\pi}\int^{2\pi}_0\frac{\rmd \alpha}{\pi}\int\rmd{\mib r}{\rm Im}g(z=0+\iu \delta).
\label{densityofstates}
\end{equation}
Substituting (\ref{analytic1}) into (\ref{densityofstates}), we obtain
\begin{equation}
N(\epsilon=0)\approx N_0 \xi_0^2/\ln\left(\xi_0/\xi_{1-}\right)\sim N_0 \xi_0^2/\ln\left(T_{\rm c}/T\right).
\label{densityofstatesestimate}
\end{equation}
This logarithmic factor ($1/\ln\left(T_{\rm c}/T\right)$) can be seen, in principle, in the specific heat at low temperatures. 
\subsection{Case II: $\Delta({\mib r},\alpha)\rightarrow \Delta_\infty e^{\iu \left(\phi+\alpha\right)}\label{subsec:CaseII}
$}

Following the method in refs.~(\citen{Eschrig}),(\citen{Kato}), the expression for $\hat g$ in the present case is given by
\begin{equation}
\hat g\sim  \frac{\pi v \exp\left[-u_+(s)\right]\hat M_+(\alpha)}{2C_+\left(z-E_+(b)\right)},
\label{analytic2}
\end{equation}
with 
$$
\hat M_+(\alpha)=\left(
\begin{array}{cc}
1,&-\iu e^{2\iu \alpha}\\
-\iu e^{-2\iu \alpha},&-1
\end{array}
\right).
$$
Here the spectral weight $\exp\left[-u_+(s)\right]$ is given by
$$
u_+(s)=\frac{2}{v}\int_0^{|s|}\rmd s'\left\{\Delta_0^+(s')+\Delta_0^-(s')\right\}.
$$
The constant $C_+$ is defined by
$$
C_+=\int_0^\infty\rmd s \exp\left[-u_+(s)\right].
$$
The roles of $u_+(s)$ and $C_+$ are similar to those of $u_-(s)$ and $C_-$ in the previous subsection. 
  
\begin{figure}
\begin{center}
\epsfysize=150pt
\epsfbox{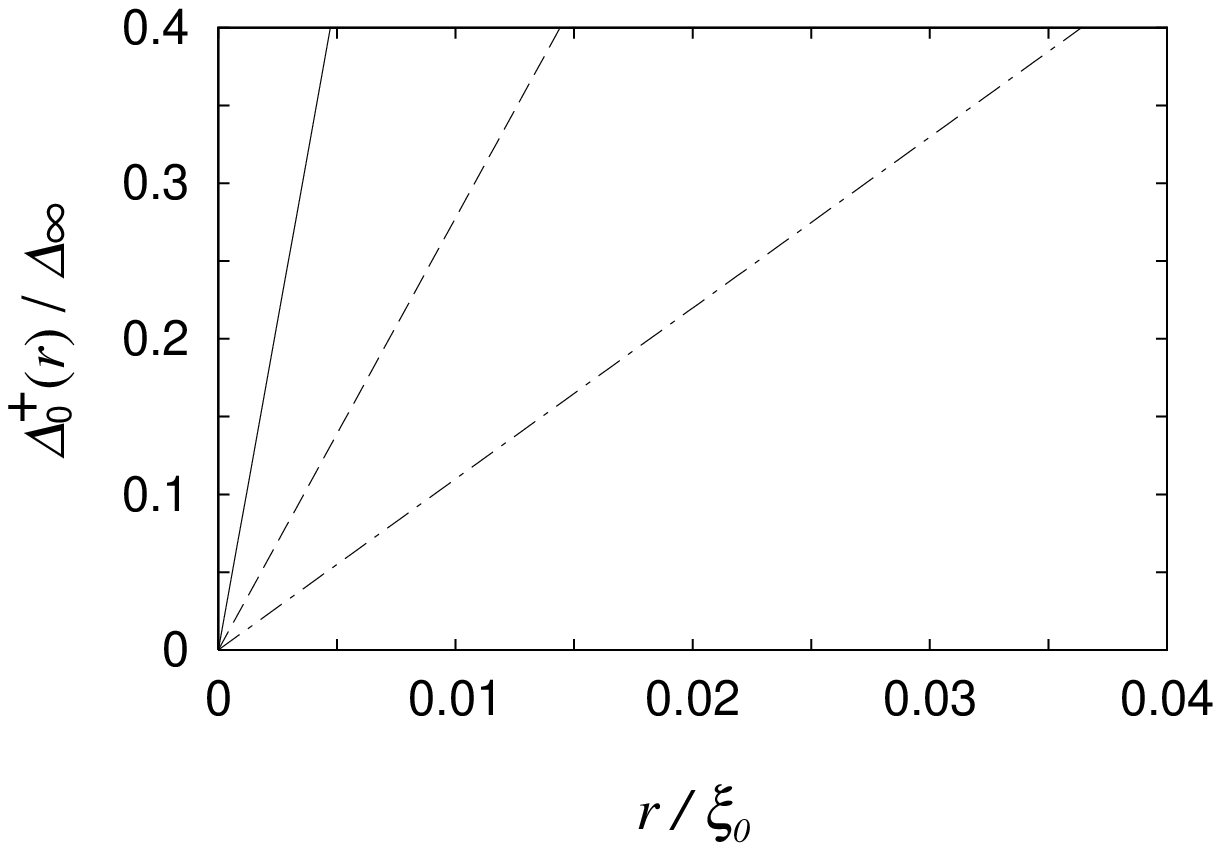}
\end{center}
\begin{center}
\epsfysize=150pt
\epsfbox{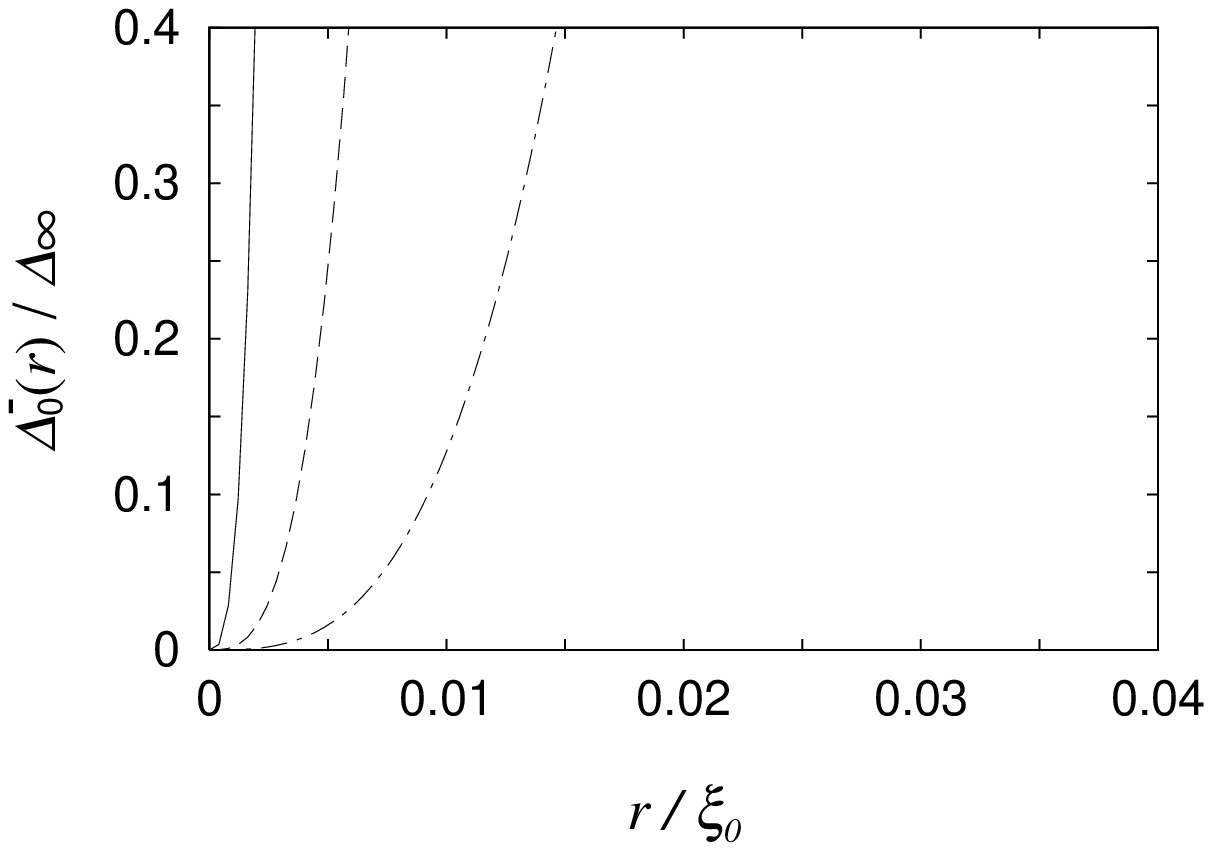}
\end{center}
\caption{Analytical results (eq.~(\ref{DeltalimitKP2}) in the text) 
for case II where total angular momentum is two. 
The initial slopes of the pair-potentials (a) $\Delta_0^+(r)$ for the dominant component 
with positive chirality and (b) the induced component $\Delta_0^-(r)$ 
with negative chirality are shown as functions of the distance $r$ 
from the vortex center, for temperatures $T/T_c=0.02$ (solid line), 
0.05 (dashed line) and 0.1 (dotted-dashed line). 
The vertical and horizontal 
axes are scaled, respectively, 
by the modulus of the bulk pair-potential $\Delta_\infty$ 
and the coherence length $\xi_0$ at zero temperature. }
\label{fig:2}
%
%
%
%
%
\end{figure}

The dispersion $E_+(b)$ of the Andreev bound state is given by
\begin{equation}
E_+(b)=\frac{b}{C_+}\int_0^\infty\frac{\rmd s}{s}\exp\left[-u_+(s)\right]\left\{\Delta_0^+(s')+3\Delta_0^-(s')\right\}.
\end{equation}
From (\ref{analytic2}) and (\ref{gapequation}), we obtain 
\begin{subeqnarray}
&\Delta^+_0(r)&=\frac{\pi v V_p}{4C_+}I_1\left(\frac{r}{\xi_{1+}}\right).\\
&\Delta^-_0(r)&=\frac{\pi v V_p}{4C_+}I_3\left(\frac{r}{\xi_{1+}}\right).
\label{DeltaKP2}
\end{subeqnarray}
Here $\xi_{1+}$ is $2T/E'_+$ with $E'_+=\rmd E_+(b)/\rmd b$. For $r\ll \xi_{1+}\equiv 2T/E_+'$, the expression (\ref{DeltaKP2}) reduces to 
\begin{subeqnarray}
&\Delta^+(r)&\rightarrow \frac{\pi v V_p}{8 C_+}\left(\frac{r}{\xi_{1+}}\right)\\ 
&\Delta^-(r)&\rightarrow \frac{\pi v V_p}{96 C_+}\left(\frac{r}{\xi_{1+}}\right)^3. 
\label{DeltalimitKP2}
\end{subeqnarray}
The result (\ref{DeltalimitKP2}) is shown in Fig.~\ref{fig:2}. The relative phase between the two components in this case is zero. If we scale the $r$-dependence in (\ref{DeltalimitKP2}) by $\xi_{1+}$, the ratio of the prefactor of each component then takes a universal value ($=12$). 
From (\ref{currentdensity}) and (\ref{analytic2}), the current density is given by ${\mib j}_+({\mib r})=j_+(r)\hat{\mib \phi}$ with 
\begin{equation}
j_+(r)=\frac{\pi e v^2 N_0I_1\left(r/\xi_{1+}\right)}{2C_+}\sim \frac{\pi e v^2 N_0}{4C_+}\left(\frac{r}{\xi_{1+}}\right). 
\end{equation} 
By an argument similar to that in the previous subsection, we find that $C_+\sim \xi_0$, 
$$
E'_+\sim \left(\Delta_\infty/\xi_0\right)\ln\left(\xi_0/\xi_{1+}\right)
$$
and
$$
\xi_{1+}/\xi_0 \sim \left(T/\Delta_\infty\right)/\ln\left(\Delta_\infty/T\right). 
$$    
Temperature dependence of the length $\xi_{1+}$ leads to the shrinkage of the distributions of the pair-potential and circular current around vortex core at $T$ much lower than $T_{\rm c}$. 
\subsection{Modification due to finite $\kappa$}\label{subsec:finitekappa}
In this subsection, we briefly discuss on the case where the GL parameter~$\kappa$ is  finite; most results in the previous subsections hold with the following modifications. 

In section~\ref{sec:Formulation}, $\epsilon$ in eqs. (\ref{qclequation}), (\ref{Riccati1}) and (\ref{ic1}) is replaced by $\tilde \epsilon=\epsilon+ev\hat{\mib p}\cdot {\mib A}({\mib r})/c$ in the case of finite $\kappa$. The  symbol $c$ denotes the velocity of light. Owing to this change, $E_-(b)$ in (\ref{Eminus}), (\ref{DeltaKP1}) and (\ref{DeltalimitKP1}) is replaced by $\tilde E_-(b)\equiv \tilde E'_- b$ with 
\begin{equation}
\tilde E'_-=E'_-+\frac{ev}{cC_-}\int_0^\infty\frac{\rmd r}{r}A(r)\exp\left[-u_-(r)\right].
\end{equation}
Here $A(r)$ is introduced through ${\mib A}({\mib r})=A(r)\left(\hat{\mib z}\times \hat{\mib r}\right)$. The length scale $\xi_{1-}$ in (\ref{DeltalimitKP1}) is replaced by $2T/\tilde E'_-$, accordingly. Similarly, $E_+(b)$ in subsection~\ref{subsec:CaseII} is replaced by $\tilde E_+(b)\equiv \tilde E'_+ b$, with
\begin{equation}
\tilde E'_+=E'_++\frac{ev}{cC_+}\int_0^\infty\frac{\rmd r}{r}A(r)\exp\left[-u_+(r)\right].
\end{equation}
The length $\xi_{1+}$ is replaced by $2T/\tilde E'_+$. 

The current density coming from the chiral branch (\ref{analytic1}) is given by ${\mib j}({\mib r})=j(r)\hat{\mib \phi}$ with 
\begin{equation}
j(r)=\frac{\pi e v^2 N_0I_1\left(r/\xi_{1-}\right)}{2C_-}\sim \frac{\pi e v^2 N_0}{4C_-}\left(\frac{r}{\xi_{1-}}\right). 
\end{equation}
As in the previous subsection, the density of states (integrated over a vortex core) at zero energy (\ref{densityofstates}) is obtained as $N_0\xi_0^2/\ln\left(T_{\rm c}/T\right)$.  
\section{Numerical Results}\label{sec:Numerical Results}     
\subsection{Case I: $\Delta({\mib r},\alpha)\rightarrow \Delta_\infty e^{\iu \left(\phi-\alpha\right)}
$}
\begin{figure}
\begin{center}
\epsfysize=150pt
\epsfbox{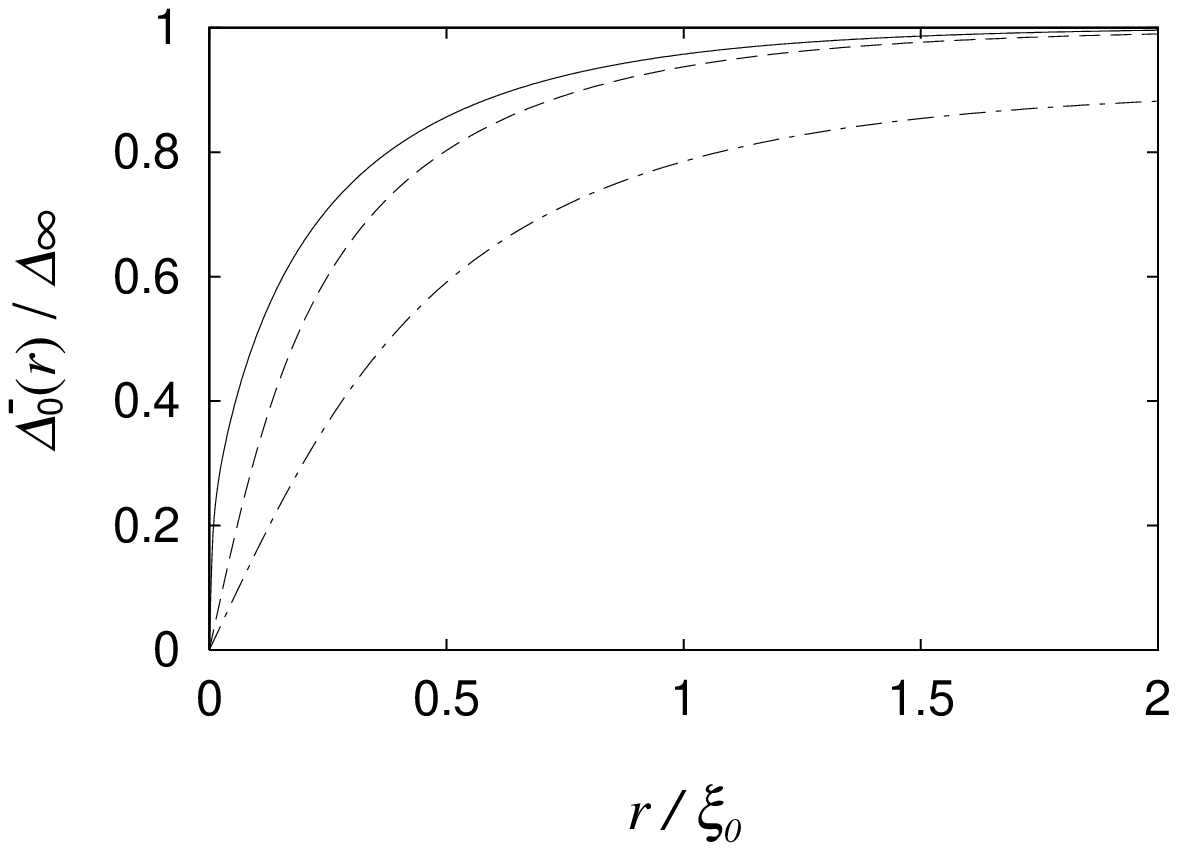}
\end{center}
\begin{center}
\epsfysize=150pt
\epsfbox{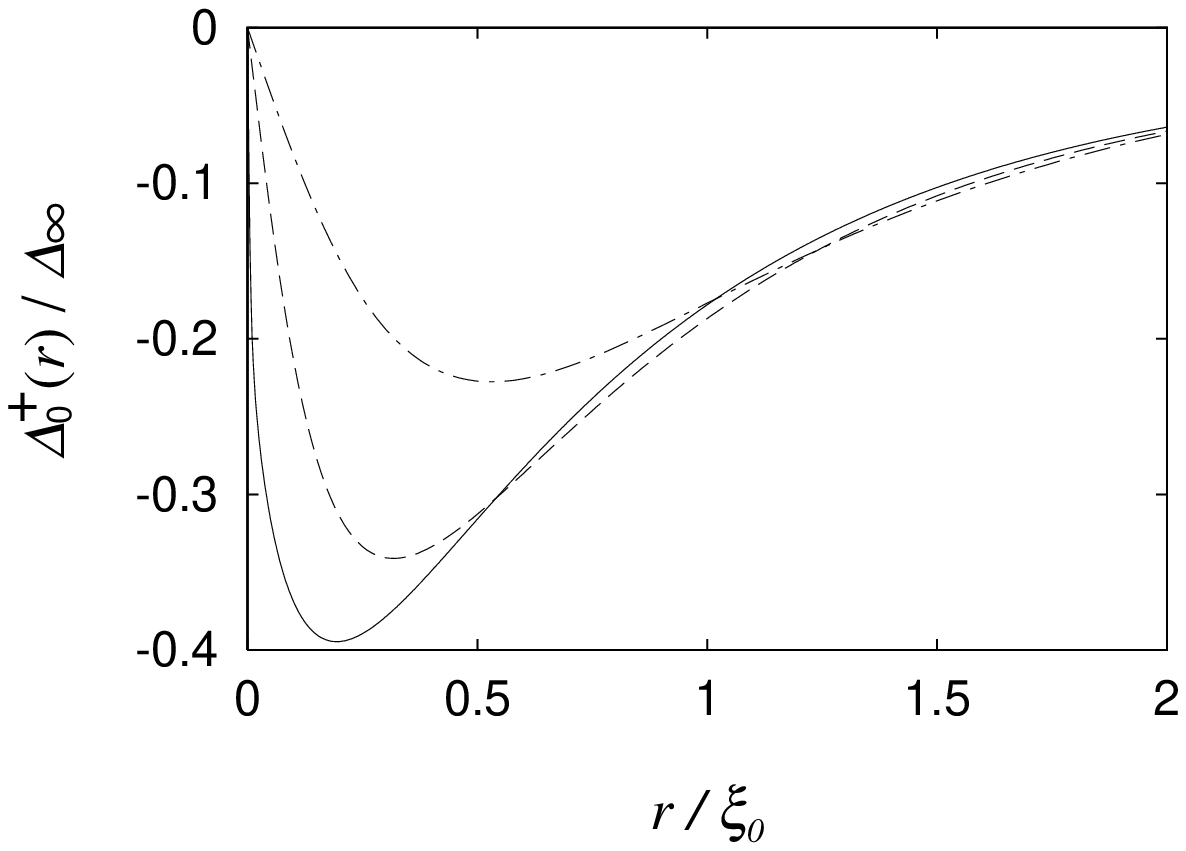}
\end{center}
\caption{Numerically determined self-consistent pair-potentials for case I where total angular momentum is zero. (a) $\Delta_0^-(r)$ of the dominant component with negative chirality and (b) the induced component $\Delta_0^+(r)$ with positive chirality are shown as functions of the distance $r$ from the vortex center, for temperatures $T/T_{\rm c}=0.02$ (solid line), 0.3 (dashed line) and 0.6 (dotted-dashed line). The vertical and horizontal axes are scaled, respectively, by the modulus of the bulk pair-potential $\Delta_\infty$ and the coherence length $\xi_0$ at zero temperature. }
\label{fig:3}
%
%
%
%
%
\end{figure}

\begin{figure}
\begin{center}
\epsfysize=150pt
\epsfbox{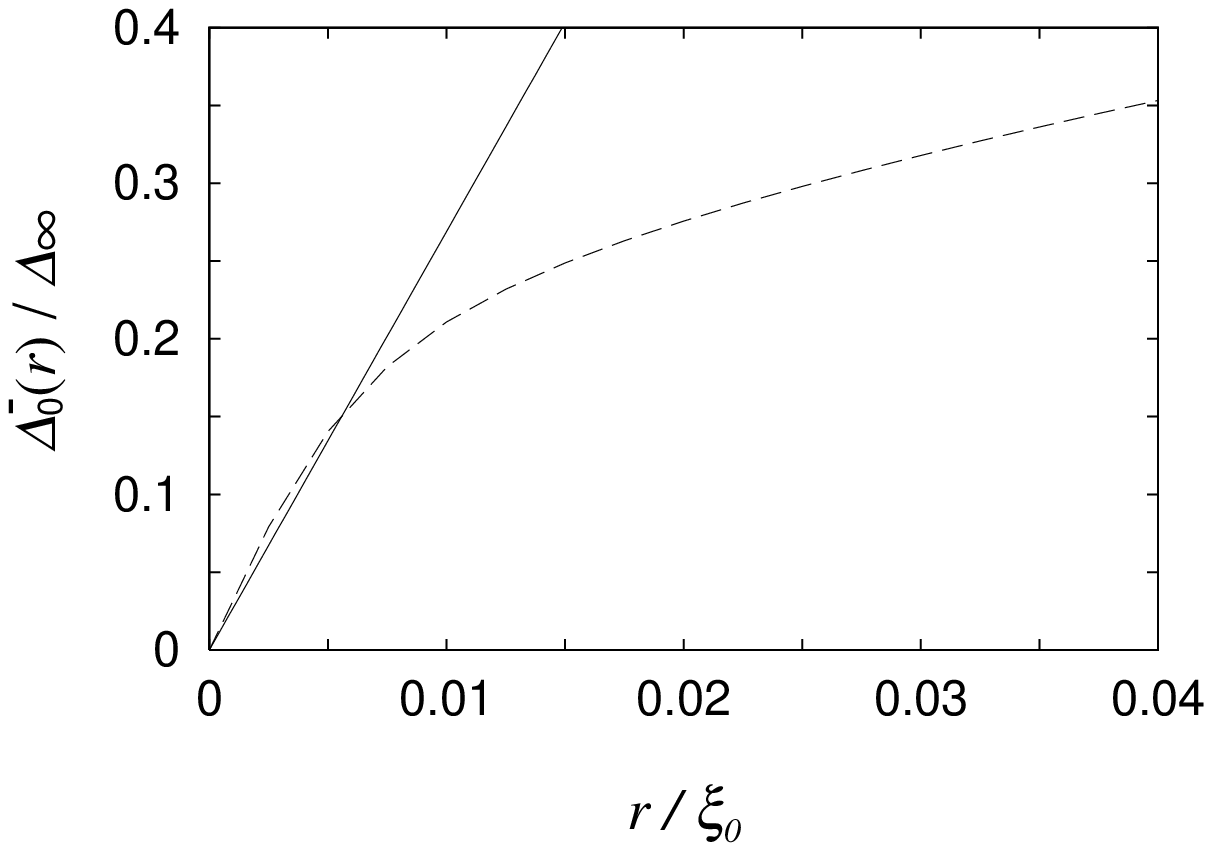}
\end{center}
\begin{center}
\epsfysize=150pt
\epsfbox{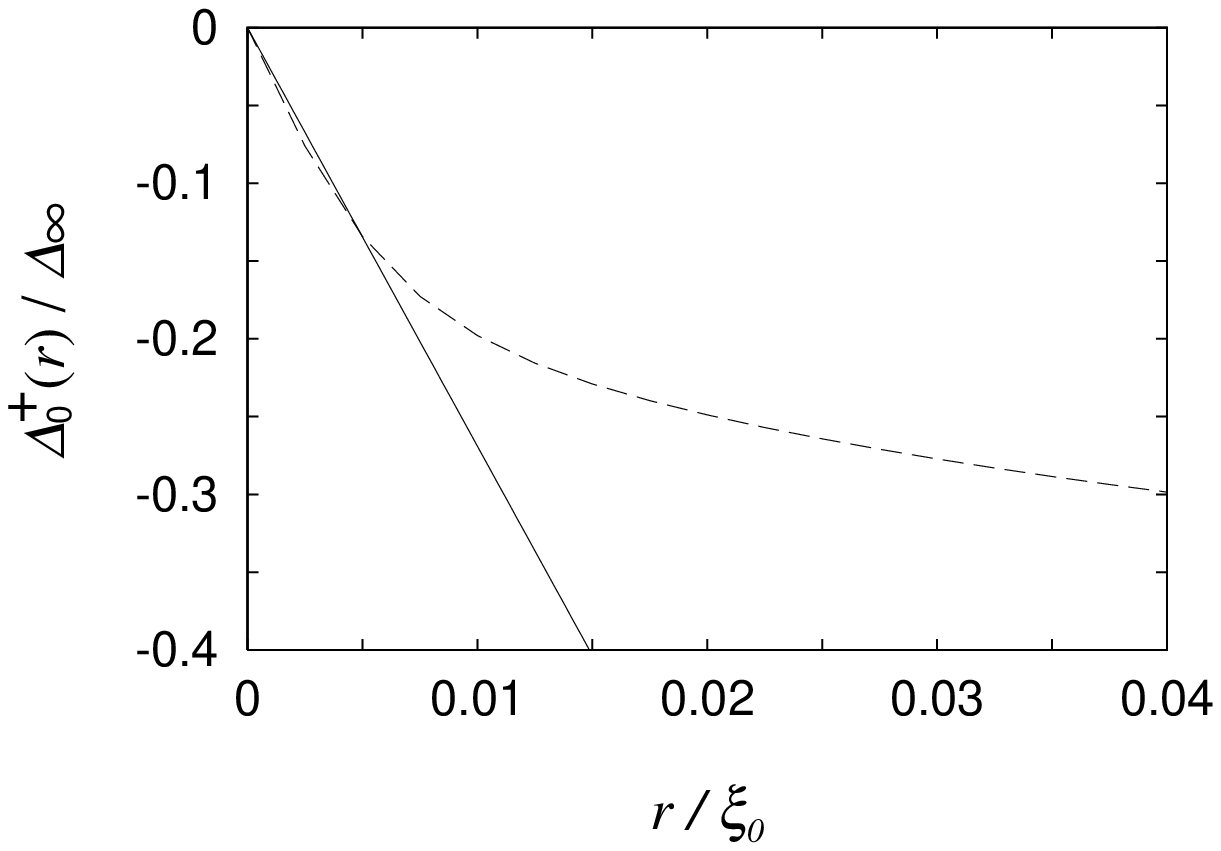}
\end{center}
\caption{Comparison between analytical (solid line) and numerical (dashed line) results on the pair-potential for case I at $T/T_c=0.02$. (a) $\Delta_0^-(r)$ of the dominant component with negative chirality and (b) the induced component $\Delta_0^+(r)$ with positive chirality are shown as functions of the distance $r$ from the vortex center. The vertical and horizontal axes are scaled, respectively, by the modulus of the bulk pair-potential $\Delta_\infty$ and the coherence length $\xi_0$ at zero temperature. In both figures, the relative deviations of analytical and numerical results are 0.12 at $r/\xi_0=0.0025$. }
\label{fig:4}
%
%
%
%
%
\end{figure}

\begin{figure}
\begin{center}
\epsfysize=150pt
\epsfbox{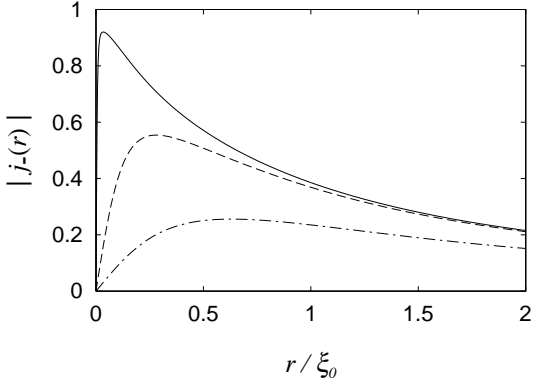}
\end{center}
\caption{Numerically determined current density is shown as a function of the distance $r$ from the vortex center for case I at temperatures $T/T_{\rm c}=0.02$ (solid line), 0.3 (dashed line) and 0.6 (dotted-dashed line). The vertical and horizontal axes are scaled, respectively, by the critical current density $j_{\rm c}=2N_0 |e|v T_{\rm c}$ in the bulk and the coherence length $\xi_0$ at zero temperature. }
\label{fig:5}
%
%
%
%
%
%
%

%
\end{figure}

\begin{figure}
\begin{center}
\epsfysize=150pt
\epsfbox{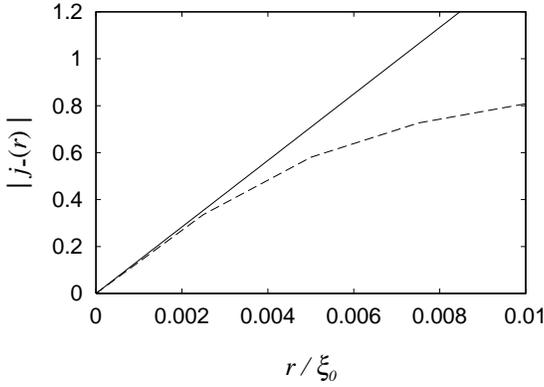}
\end{center}
\caption{Comparison between analytical (solid line) and numerical (dashed line) results on the current density for case I at $T/T_c=0.02$. 
The vertical and horizontal axes are scaled, respectively, by the critical current density $j_c=2N_0 |e|v T_{\rm c}$ in the bulk and the coherence length $\xi_0$. }
\label{fig:6}
%
%
%
%
%
%
\end{figure}

Figure~\ref{fig:3} shows the numerically determined self-consistent pair-potentials as a function of $r/\xi_0$, at temperatures $T/T_{\rm c}=0.02$ (solid line), 0.3 (dashed line) and 0.6 (dotted-dashed line). The relative phase between the dominant ($\Delta^-_0(r)$) and the induced ($\Delta^+_0(r)$) components is $\pi$. Further, the spatial variations near the vortex center becomes steeper with decreasing temperature and hence the shrinking of the vortex core at low temperatures is confirmed numerically in the chiral p-wave superconductors. These two points have been expected from the analytical results in the previous section. We compare the numerical results at $T/T_{\rm c}=0.02$ with the analytical result eq. (\ref{DeltalimitKP1}) in fig.~\ref{fig:4}. For both components of the pair-potential, analytical results explain the numerical results within the accuracy of 12\% at $r/\xi_0=0.0025$. 
Figure~\ref{fig:5} shows numerical results on the circular current density surrounding the vortex core for temperatures $T/T_c=0.02$ (solid line), 0.3 (dashed line) and 0.6 (dotted-dashed line). The current density and the distance from the vortex center are, respectively, scaled by the critical current density $j_{\rm c}=2N_0 |e|v T_{\rm c}$ in the bulk and the coherence length $\xi_0$ at zero temperature. We note that the $\hat\phi$ component is always negative, as expected from the analytical calculations. We can see in the figure that the distribution of the circular current becomes more concentrated with decreasing temperature. This feature is consistent with the results on the pair-potential, which shows the shrinkage of the core at low temperatures. 
Figure~\ref{fig:6} blows up the vicinity of the core center at $T/T_{\rm c}=0.02$. The numerical result is shown by the dashed line. The analytical result (solid line) is displayed for comparison. Obviously, the numerical result approaches the analytical result asymptotically near the vortex core. 
\subsection{Case II: $\Delta({\mib r},\alpha)\rightarrow \Delta_\infty e^{\iu \left(\phi+\alpha\right)}
$}
\begin{figure}
\begin{center}
\epsfysize=150pt
\epsfbox{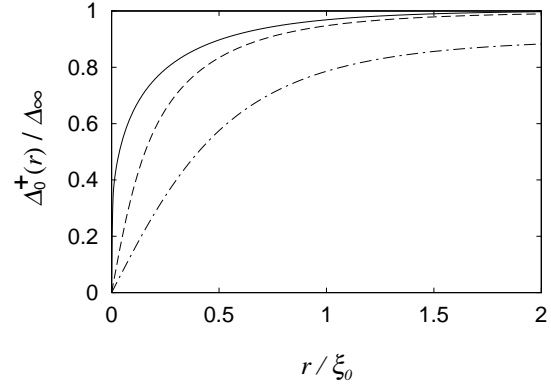}
\end{center}
\begin{center}
\epsfysize=150pt
\epsfbox{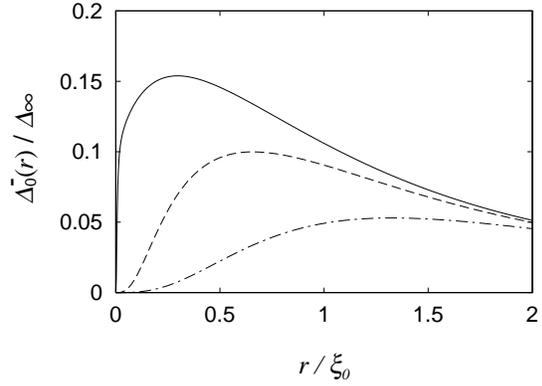}
\end{center}
\caption{Numerically determined self-consistent pair-potentials 
for case II where total angular momentum is two. 
The pair-potentials (a) $\Delta_0^+(r)$ of the dominant component 
with positive chirality and (b) the induced component $\Delta_0^-(r)$ 
with negative chirality are shown as functions of the distance $r$ 
from the vortex center, for temperatures $T/T_c=0.02$ (solid line), 
0.3 (dashed line) and 0.6 (dotted-dashed line). 
The vertical and horizontal 
axes are scaled, respectively, 
by the modulus of the bulk pair-potential $\Delta_\infty$ 
and the coherence length $\xi_0$ at zero temperature. }
\label{fig:7}
%
%
%
%
%
\end{figure}

\begin{figure}
\begin{center}
\epsfysize=150pt
\epsfbox{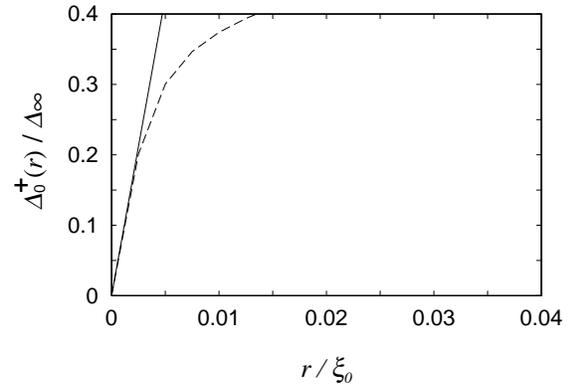}
\end{center}
\caption{Comparison between analytical (solid line) and numerical (dashed line) results on the pair-potential $\Delta_0^+(r)$ for the dominant component with positive chirality for case II at $T/T_c=0.02$. 
The vertical and horizontal axes are scaled, respectively, by the modulus of the bulk pair-potential $\Delta_\infty$ and the coherence length $\xi_0$ at zero temperature. }
\label{fig:8}
%
%
%
%
%
\end{figure}

Figure~\ref{fig:7} shows numerically the determined self-consistent pair-potentials at temperatures $T/T_{\rm c} =0.02$ (solid line), 
0.3 (dashed line) and 0.6 (dotted-dashed line). Once again, we see that the core shrinking occurs even in this case. The relative phase between the two components of the pair-potential is 0 in this case. These two points have been anticipated from the analytical calculation. For the induced component $\Delta_0^-(r)$, the spatial variation is expected to be proportional to $(r/\xi_0)^3$ near the core at low temperatures. Correspondingly, figure~\ref{fig:7}(b) shows the positive curvature near the core. Precisely speaking, we have confirmed that $\Delta_0^-(r)$ at $T/T_{\rm c}=0.1$ is proportional to $(r/\xi_0)^3$ in the range $r/\xi_0<0.01$. At the lowest temperature $T/T_{\rm c}=0.02$, however, the cubic dependence on $r/\xi_0$ cannot be confirmed at the distance larger than $r/\xi_0=0.002$ within our numerical accuracy. Thus, a comparison with analytical result is made only for the dominant component. 
Figure~\ref{fig:8} shows the analytical (solid line) and the numerical (dashed line) results on the pair-potential for the dominant component at $T/T_c=0.02$. We find good agreement between the two results for distances smaller than $r/\xi_0=0.0025$. 

\begin{figure}
\begin{center}
\epsfysize=150pt
\epsfbox{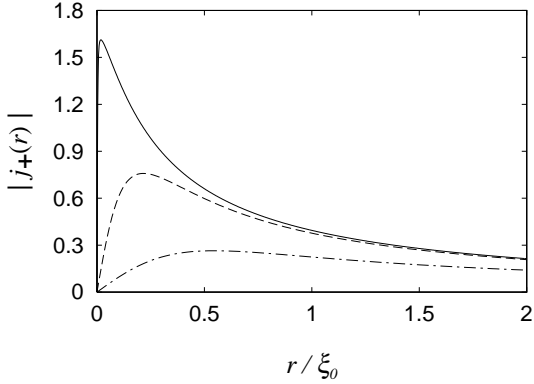}
\end{center}
\caption{Numerically determined current density is shown as a function of the distance $r$ from the vortex center for case II at temperatures $T/T_c=0.02$ (solid line), 0.3 (dashed line) and 0.6 (dotted-dashed line). The vertical and horizontal axes are scaled, respectively, by the critical current density $j_{\rm c}=2N_0 |e|v T_{\rm c}$ in the bulk and the coherence length $\xi_0$ at zero temperature. }
\label{fig:9}
%
%
%
%
%
%
\end{figure}

Now we turn to the circular current. In fig.~\ref{fig:9}, the spatial distribution of the circular current is shown for $T/T_c=0.02$ (solid line), 0.3 (dashed line) and 0.6 (dotted-dashed line). The unit of the current density is again $j_{\rm c}=2N_0 |e|v T_{\rm c}$. Overall features are similar with fig.~\ref{fig:5}; narrowing of the distribution at low temperatures are equally observed in both cases. The magnitude of current density near the core in the present case is, however, about 1.5 times larger than in case I. This means that the cost of the kinetic energy of the condensate due to the introduction of vortex is larger in case II than that in case I. This energy difference favors case I against case II as a stable state in the presence of magnetic field. This observation is consistent with the Ginzburg-Landau theory for chiral p-wave superconductors\cite{Heeb}. 

\begin{figure}
\begin{center}
\epsfysize=150pt
\epsfbox{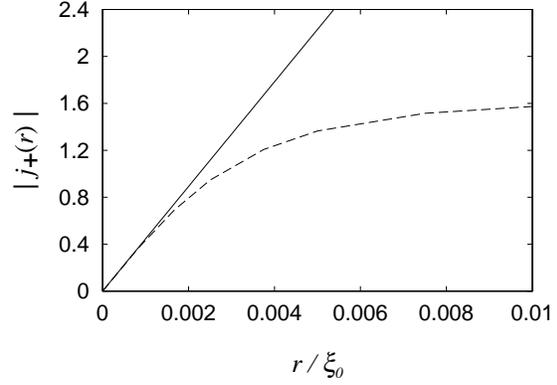}
\end{center}
\caption{Comparison between analytical (solid line) and numerical (dashed line) results on the current density for case II at $T/T_c=0.02$. 
The vertical and horizontal axes are scaled, respectively, by the critical current density $j_{\rm c}=2N_0 |e|v T_{\rm c}$ in the bulk and the coherence length $\xi_0$ at zero temperature.}
\label{fig:10}
%
%
%
%
%
\end{figure}

In fig.~\ref{fig:10}, we compare the numerical result (dashed line) with the analytical one (solid line) at $T/T_c=0.02$. The numerical result approaches asymptotically and collapses on the analytical result at $r/\xi_0\sim 0.001$. 
\section{Discussion}\label{sec:Discussion}
Most aspects of the shrinkage of vortex core in chiral superconductors are found to be similar to those in the isotropic s-wave superconductors. However, there are two noticeable differences between the two cases, which is discussed below. 

First, we discuss the stability of the vortex shrinkage against the impurities. In refs.~\citen{Volovik,Ivanov,Bocquet,Kato}, it was shown that the impurity effects within vortex core in chiral superconductors are quite different from those in the isotropic superconductors. In the isotropic s-wave superconductors, the impurity scattering rate $\Gamma(\epsilon,T)$ for the chiral branch with energy $\epsilon$ at temperature $T$ was estimated to be $\Gamma_{\rm n}\ln\left(T_{\rm c}/\epsilon\right)/\ln\left(T_{\rm c}/T\right)$\cite{LO76,KopninLopatin} in the Born limit, where $\Gamma_{\rm n}$ is the scattering rate in the normal state. The scattering rates $\Gamma(\epsilon,T)$ for chiral superconductors can be estimated by following the method used in ref.~\citen{Kato}. In case I, $\Gamma(\epsilon,T)/\Gamma_{\rm n}$ turns out to be at most ${\cal O}\left(\epsilon/T_{\rm c}\right)$, while $\Gamma(\epsilon,T)/\Gamma_{\rm n}\sim 1/\ln\left(T_{\rm c}/T\right)$ in case II. We note that the results in ref.~\citen{Kato} on $\Gamma$ still hold in the present case although the induced (minor ) component has been neglected in ref.~\citen{Kato}. In the presence of impurities, the length $\xi_{1\pm}$ should be replaced by 
$$
\xi_{1\pm}={\rm Max}\left[T,\Gamma\right]/E'_{\pm}(T).
$$ 
Thus $\xi_{1\pm}$ becomes temperature-independent at $T(<\Gamma)$ and hence the vortex ceases to shrink. Within the logarithmic accuracy, $\Gamma\sim \Gamma_{\rm n}$ in the isotropic s-wave and chiral p-wave (Case II) superconductors. In the two cases, therefore, the temperature range where the KP effect occurs is given by $\Gamma_{\rm n}<T\ll T_{\rm c}$. In the chiral p-wave (Case I) superconductors, on the other hand, $\Gamma$ is smaller than $T$ under the condition $\Gamma_{\rm n}/T_{\rm c}\ll 1$ (clean superconductors). Thus, in case I, the presence of impurity imposes no restrictions on the temperature range where the KP effect occurs. The importance of this fact will be understood by recalling the current experimental situations on the KP effect~\cite{Sonier}. The $\mu$ SR experiments on NbSe$_2$ with $T_{\rm c}\approx 7.1$K show that the vortex core size ceases to shrink at $T\approx 1$K and the temperature dependence of the core size at $T>1$K is much weaker than theoretical prediction~\cite{Miller}. The saturation $T\approx 1$K is presumably due to the impurity scattering, which smears out the structure of the energy spectrum of the bound states inside core. Takagi {\it et al}~\cite{Takagi} measured the flux flow conductivity of NbSe$_2$ with high purity ($RRR=100$) and did not observe the logarithmic temperature dependence, which could be regarded as an evidence for the KP effect.  Therefore, if we try to observe the KP effect in NbSe$_2$, we have to prepare extremely pure samples with RRR $\gg 100$. Instead, chiral p-wave superconductors give us the opportunity to see the KP effect, by suppressing the impurity scattering rate in vortex core. 

Second, we make a remark on the local recovery of the symmetry in the pair-potential in case I $(L_z=0)$. In this case, the pair-potential (\ref{Deltaralpha}) with (\ref{DeltalimitKP1}) is invariant under the combined action of time-reversal operation 
$(\Delta({\mib r},\alpha)\rightarrow -\Delta^*({\mib r},\alpha+\pi))$ (minus sign exists for the odd-parity case~\cite{SigristUeda}) and a gauge transformation $(\Delta({\mib r},\alpha)\rightarrow -\Delta({\mib r},\alpha))$. The local recovery of symmetry does not occur within the core in case II 
$(L_z=2)$. Thus, {\it the cancellation of angular momentum} due to vorticity and chirality is crucial for this phenomenon. It should be noted that such a restoration of symmetry occurs only in the pair-potential; for example, there exists non-vanishing circular current within the core. However, this {\it pseudo} time-reversal-symmetry gives an intuitive interpretation on the impurity effects in vortex core. From the symmetry and angular momentum, the chiral p-wave vortex core with $L_z=0$ can be identified with the s-wave superconductors {\it in the absence of the magnetic field}. For the latter, non-magnetic impurities are known to be harmless (the Anderson theorem~\cite{Anderson}). These two explain why non-magnetic impurities are harmless in the chiral p-wave vortex with $L_z=0$. 
         
\section{Conclusion}\label{sec:Conclusion}
The physics of isolated vortices in pure chiral superconductors is similar to that in pure isotropic s-wave superconductors. The presence of impurities little affects the chiral branch in the chiral superconductors. Therefore, it is expected that the KP effect in the chiral superconductors is much more accessible in experiments than that in s-wave superconductors. Further, the local recovery of modified time-reversal symmetry around vortex core is found when the vorticity and chirality have the opposite sign.   
\acknowledgement
We thank M.~Matsumoto and M.~Sigrist for their correspondence. Y.K. also thank A.~Tanaka for his critical reading of the manuscript and T.~Hanaguri, A.~Maeda, K.~Takagi and Y.~Tsuchiya for their useful discussions. 
This work is partly supported by Grant-in-Aid for Scientific Research on Priority Areas (A) of ^^ ^^ Novel Quantum Phenomena in Transition Metal Oxides" (12046225) from the Ministry of Education, Science, Sports and Culture and Grant-in-Aid for Encouragement of Young Scientists from Japan Society for the Promotion of Science (12740203). Part of the numerical calculations was processed by utilizing the communal facilities of the Computer Center, Okayama University.


\begin{thebibliography}{99}
\bibitem{Maeno}Y. Maeno, H. Hashimoto, K. Yoshida, S. Nishizaki, T. Fujita, J.~G.~Bednorz and F. Lichtenberg: Nature {\bf 372} (1994) 532.  
\bibitem{Ishida}K.~Ishida, H.~Mukuda, Y.~Kitaoka, K.~Asayama, Z.~Q.~Mao, Y.~Mori and Y. Maeno: Nature {\bf 396} (1998) 658. 
\bibitem{Luke}G.~M.~Luke, Y.~Fudamoto, K.~M.~Kojima, M.~I.~Larkin, J.~Merrin, B.~Nachumi, Y.~J.~Uemura, Y.~Maeno, Z.~Q.~Mao, Y.~Mori, H.~Nakamura, M.~Sigrist: Nature {\bf 394} (1998) 558. 
\bibitem{Sigrist}M.~Sigrist, D.~Agterberg, A.~Furusaki, C.~Honerkamp, K.~K.~Ng, T.~M.~Rice and M.~E.~Zhitomirsky: Physica C {\bf 317-318} (1999) 134. 

\bibitem{Volovik} G. E. Volovik: Pis'ma Zh. Eksp. Teor. Fiz. {\bf 70} (1999) 601 [JETP Lett. {\bf 70} (1999) 609]. 
\bibitem{Bocquet}M. Bocquet, D. Serban and M. Zirnbauer: Nucl. Phys. B 578 (2000) 628 (cond-mat/9910480).
\bibitem{Ivanov} D. A. Ivanov: cond-mat/9911147.
\bibitem{Kato}Y. Kato: J. Phys. Soc. Jpn {\bf 69} (2000) 3378, (cond-mat/0009296).
%
%
\bibitem{KramerPesch}L. Kramer and W. Pesch: Z. Phys. {\bf 269} (1974) 59; W. Pesch and L. Kramer: J. Low Temp. Phys. {\bf 15} (1974) 367.
\bibitem{Gygi}F. Gygi and M. Schl\" uter: Phys. Rev. B{\bf 43} (1991) 7609. 
\bibitem{Volovik2}G. E. Volovik: Pis'ma Zh. Eksp. Teor. Fiz. {\bf 58} (1993) 444 [JETP Lett. {\bf 58} (1993) 455].
\bibitem{Ichioka}M. Ichioka, N. Hayashi, N. Enomoto and K. Machida: Phys. Rev. B {\bf 53} (1996) 15316. 
\bibitem{Sonier}J~.E.~Sonier, J.~H.~Brewer and R.~F.~Kiefl: Rev. Mod. Phys. {\bf 72} (2000) 769. 
\bibitem{HeebThesis}One of the authors (Y.K.) was informed from M. Sigrist that R. Heeb studied the Kramer-Pesch effect in the chiral p-wave superconductors within the chiral p-wave vortex (Doctor Thesis, ETH Z\"urich 2000).
\bibitem{M&H}M. Matsumoto and M. Sigrist:
 J. Phys. Soc. Jpn. {\bf 68} (1999) 724;
M. Matsumoto and M. Sigirst: Physica B {\bf 281-282} (2000) 973; 
M. Matsumoto and R. Heeb: cond-mat/0101155.
%

\bibitem{Takigawa}M. Takigawa, M. Ichioka, K. Machida and M. Sigrist: cond-mat/0106367. 
\bibitem{KM}M. Kato and K. Maki: Prog. Theor. Phys. {\bf 103} (2000) 867. 
\bibitem{MKM}Y. Morita, M. Kohmoto and K. Maki: Phys. Rev. Lett. {\bf 78} (1997) 4841; {\bf 79} (1997) 4514; Europhys. Lett. {\bf 40} (1997) 207; Int. J. Mod. Phys. B {\bf 12} (1998) 989.
\bibitem{HIIM}N. Hayashi, T. Isoshima, M. Ichioka and  K. Machida: Phys. Rev. Lett. {\bf 80} (1998) 2921. 
\bibitem{F&T}M. Franz and Z. Te\v sanovi\'c: Phys. Rev. Lett. {\bf 80} (1998) 4763.  

\bibitem{Akima}T. Akima, S. NishiZaki and Y. Maeno: J. Phys. Soc. Jpn. {\bf 68} (1999) 694. 
%
\bibitem{Eilenberger}G. Eilenberger: Z. Phys. {\bf 214} (1968) 195. 
\bibitem{LO68}A. Larkin and Yu. Ovchinnikov: Zh. Eksp. Teor. Fiz. {\bf 55} (1968) 2262 [Sov. Phys. JETP {\bf 28} (1969) 1200]. 
\bibitem{Eliashberg}G. M. Eliashberg: Zh. Eksp. Teor. Fiz. {\bf 61} (1971) 1254 [Sov. Phys. JETP {\bf 34} (1972) 668].
\bibitem{SereneRainer}J. W. Serene and D. Rainer: Phys. Reports {\bf 101} (1983) 221. 

\bibitem{Matsumotodomain}M. Matsumoto and M. Sigrist: J. Phys. Soc. Jpn. {\bf 68} (1999) 994; Physica B {\bf 284-288} (2000) 545. 


\bibitem{Andreev}A. F. Andreev: Zh. Eksp. Teor. Fiz. {\bf 46} (1964) 1823 [Sov. Phys. JETP {\bf 19} (1964) 1228].    

%
%
%
%
%
%
%
%
%
%
\bibitem{Nagato1}Y. Nagato, K. Nagai and J. Hara: J. Low Temp. Phys. {\bf 93} (1993) 33.
\bibitem{Higashitani}S. Higashitani and K. Nagai: J. Phys. Soc. Jpn. {\bf 64} (1995) 549. 
\bibitem{Nagato2}Y. Nagato, S. Higashitani, K. Yamada and K. Nagai: J. Low Temp. Phys. {\bf 103} (1996) 1. 
\bibitem{SchopohlMaki} N. Schopohl and K. Maki: Phys. Rev. B {\bf 52} (1995) 490. 
\bibitem{Schopohl}N. Schopohl: in {\it Quasiclassical Methods in Superconductivity \& Superfluidity } edited by D. Rainer and J. Sauls, (1996) 88; K. Nagai: {\it ibid} (1996) 198.
\bibitem{Eschrig}M. Eschrig: Ph. D Thesis, University of Bayreuth (1997). 
\bibitem{Recipe}W. H. Press, S. A. Teukolsky, W. T. Vettering and B. P. Flannert, {\it Numerical Recipes in FORTRAN, }2nd ed. (Cambridge Univ. Press, London, 1992), p. 708. 

%
%
%
%
%
%
%
%

\bibitem{Atiyah}M. Atiyah, V. Patodi and I. Singer: Camb. Phil. Soc. {\bf 77} (1975) 43. 

\bibitem{Caroli}C. Caroli, P. G. de Gennes and J. Matricon: Phys. Lett. {\bf 9} (1964) 307. 
\bibitem{Heeb}R. Heeb and D. F. Agterberg: Phys. Rev. B {\bf 59} (1999) 7076.
%

\bibitem{LO76}A. Larkin and Yu. Ovchinnikov: Pis'ma Zh. Eksp. Teor. Fiz. {\bf 23} (1976) 210 [Sov. Phys. JETP Lett. {\bf 23} (1976) 187].
%
%
\bibitem{KopninLopatin}N. B. Kopnin and A. V. Lopatin: Phys. Rev. B {\bf 51} (1995) 15291.
\bibitem{Miller}R. Miller, R. F. Kiefl, J. H. Brewer, J. Chalhalian, S. Dunsiger, G. D. Morris, J. E. Sonier and W. A. MacFarlane: Phys. Rev. Lett. {\bf 85} (2000) 1540. 

\bibitem{Takagi}K.~Takagi {\it et al.} unpublished.
\bibitem{SigristUeda}M. Sigrist and K. Ueda: Rev. Mod. Phys. {\bf 63} (1991) 239. 
\bibitem{Anderson}P. W. Anderson: J. Phys. Chem. Sol. {\bf 11} (1959) 26. 
\end{thebibliography}
\end{document}